\documentclass[11pt]{article}
\usepackage{graphicx}
\usepackage{latexsym}
\def\a{\alpha'}

\def\CE{{\cal E}}
\def\CL{{\cal L}}

\def\CF{{\cal F}}

\def\CO{{\cal O}}

\def\centeron#1#2{{\setbox0=\hbox{#1}\setbox1=\hbox{#2}\ifdim
   \wd1>\wd0\kern.48\wd1\kern-.48\wd0\fi
   \copy0\kern-.48\wd0\kern-.48\wd1\copy1\ifdim\wd0>\wd1
   \kern.48\wd0\kern-.48\wd1\fi}}

\newcommand{\beq}{\begin{equation}}
\newcommand{\eeq}{\end{equation}}
\newcommand{\bea}{\begin{eqnarray}}
\newcommand{\eea}{\end{eqnarray}}
\newcommand{\ba}{\begin{array}}
\newcommand{\ea}{\end{array}}

\newcommand{\p}{\partial}
\newcommand{\nn}{\nonumber}

\newcommand{\half}{\frac{1}{2}}

\newcommand{\bfmu}{{\mbox{\boldmath $\mu$}}}

\renewcommand{\thefootnote}{\fnsymbol{footnote}}
\begin{document}

\hskip3cm

\centerline{\LARGE \bf Two-dimensional Black Holes}
\vskip0.5cm \centerline{\LARGE \bf in a Higher Derivative Gravity
 and Matrix Model}

\vskip2cm

\centerline{\Large Kwangho Hur\footnote{hurxx018@physics.umn.edu}, Seungjoon Hyun\footnote{ hyun@phya.yonsei.ac.kr}, Hongbin Kim\footnote{hongbin@yonsei.ac.kr},
Sang-Heon Yi\footnote{shyi@phya.yonsei.ac.kr}}

\hskip2cm

\begin{quote}
Department of Physics, College of Science, Yonsei University,
Seoul 120-749, Korea
\end{quote}

\hskip2cm


\vskip2cm

\centerline{\bf Abstract}
We construct perturbatively a class of charged black hole solutions in type 0A string theory with higher derivative terms. They have extremal limit, where the solution interpolates smoothly between near horizon $AdS_2$ geometry and the asymptotic linear dilaton geometry. We compute the free energy and the entropy of those solution using various methods. In particular, we show that there is no correction in the leading term of the free energy in the large charge limit. This supports the duality of the type 0A strings on the extremal black hole and the 0A matrix model in which the tree level free energy is exact without any $\a$ corrections.
\thispagestyle{empty}
\renewcommand{\thefootnote}{\arabic{footnote}}
\setcounter{footnote}{0}
\newpage

\section{Introduction}
Black holes are interesting objects in gravity and string theory,
which may be awaiting the complete understanding on the quantum
nature of gravity. Since black holes have a large curvature
value region near the singularity, the non-perturbative formulation
of gravity theory or its non-perturbative stringy generalization
may be needed to understand the full quantum nature of black holes. Though full non-perturbative descriptions of M/string theories are not available yet, there are several interesting toy string models whose nonperturbative descriptions are known.   One such class of toy models is the one of matrix models which correspond to the non-perturbative formulation of noncritical string  theories. Specifically, various two
dimensional string theories on the flat background can be
reformulated in terms of matrix models. In some cases, the dual
matrix models are believed to be a complete non-perturbative
formulation of the corresponding noncritical string theories. Then, one may naturally ask whether there are black hole solutions in type 0A string theory, and if there are,  how they can be incorporated and understood in the context of dual  via matrix models.

It was proposed in~\cite{Gukov:2003yp} that the type 0A matrix
model with $\mu=0$ and non-zero RR fluxes $q_+=q_-=q$ is dual to 0A string
theory on the extremal black hole~\cite{Berkovits:2001tg}. Later on, the type 0A matrix model was generalized~\cite{Maldacena:2005he} to incorporate two
kinds of RR fluxes and found that it depends only on the combined
flux\footnote{There is an additional term which depends on the
difference between RR fluxes, but it was irrelevant to arguments
for black holes~\cite{Maldacena:2005he}.} $Q=q_{+}+q_{-}$.
Since the generalized  matrix model is the
same as the matrix model with just one kind of RR
flux and there is no evidence for
the existence of black holes in the type 0A matrix model side, it was argued that the matrix model is not related to the 0A strings on black
holes~\cite{Maldacena:2005he}. Furthermore, the curvature radius of the black hole solution is the order of string length scale, independent of charge, and therefore the low energy gravity can not be trusted and it is not clear whether the black hole exists at all. 
Nevertheless, there are some pursuits of matching between the matrix
model and type 0A strings on black holes or AdS
space~\cite{Strominger:2003tm}\cite{Aharony:2005hm}\cite{Horava:2007ds}
with partial success.

In this paper we will try to extend these efforts by including lowest
order $\a$ corrections in the low energy effective gravity for type
0A string theory. As we mentioned, the curvature radius of the black hole is the order of string length scale, and therefore higher derivative terms should be taken into account. One of the motivation of this work is to determine how the behavior of the   black hole geometry is modified under the higher derivative correction.

We find the perturbative evidence of the existence of charged black holes even with the higher derivative terms in the type 0A string theory, which interpolate smoothly between the near horizon geometry and the asymptotic geometry. In the case of the extremal black hole, the near horizon geometry is $AdS_2$, as usual. This is in contrast to the four dimensional cases where it is not easy to find the interpolating solutions.
We compute the exact entropy of this extremal black hole using Sen's formalism and find the condition on the coefficient $a$ of the higher derivative term in order to have the extremal black hole solution. We also compute the free energy of the non-extremal black holes using Euclidean action approach and Wald's Noether charge method. In the extremal limit, the leading term, in the large charge limit, of the free energy turns out to be unaffected  under the $\a$-correction. This agrees with the result from the matrix model, which supports the duality of those two models.

The organization of the paper is as follows.
In section 2, we briefly review some relevant aspects on the type 0A string theory and its dual 0A matrix model. We also review the black hole solutions in type 0A string theory.
In section 3, we construct charged black hole solutions in the presence of higher derivative terms in the metric. We construct solutions, perturbatively in the coefficient $a$, and find the black hole geometries which interpolate near horizon region and asymptotic region, which is linear dilaton geometry. In section 4, we compute the free energy and the entropy of the charged black hole solutions, for both extremal and non-extremal cases using various approaches. The computation supports the duality between the 0A matrix model and the 0A strings on the extremal black hole geometry.
In section 5, we draw some conclusions.   In appendix A, we review some relevant results in the Noether charge method and in section B, we derive a generic relation between the temperature dependence of the radius of the time-like Killing horizon for a non-extremal black hole and the curvature radius of the near-horizon geometry of the corresponding extremal black hole.

\section{Black holes in the 0A string theory and the 0A matrix model}

The nonchiral projection of two dimensional fermionic string
theories gives rise to two type 0 string theories, so called type
0A and type 0B. Only NS-NS and RR sectors survive under the
projection, and thus the type 0 theories contain bosonic fields
only. In type 0A string theory, the NS-NS sector contains a
graviton, a dilaton and a tachyon, while the RR sector includes
two one-form gauge
fields~\cite{Douglas:2003up}\cite{Takayanagi:2003sm}. It was known
that the low energy effective theory of the 0A strings admits
charged black hole solutions~\cite{Berkovits:2001tg}. Based on the
computation of the free energy, it was argued that the 0A string
theory on those black holes is dual to the 0A matrix
model~\cite{Gukov:2003yp}\cite{Danielsson:2004xf}\cite{Olsson:2005en}.
In the first part of this section, we review the low energy
effective theory of 0A strings and its black hole solutions. In
the second part, we give some relevant features in the 0A matrix
model for the comparison with black hole side results.

\subsection{Black holes at the lowest order in $\a$ : Review}

The low energy effective action of type 0A string theory at the
lowest order in $\a$ is given by~\cite{Douglas:2003up}
\bea I_{0} &=& \int d^2 x \sqrt{-g}\,\bigg[\frac{1}{2\kappa^2}~
e^{-2\Phi}\Big(R + 4\nabla_{\mu}\Phi\nabla^{\mu}\Phi +
\frac{8}{\a} -f_{1}(T)(\nabla T)^2 + f_{2}(T)\Big) \nn \\
&&\qquad \qquad \qquad  -\frac{2\pi\a}{4}f_3(T)(F^{+})^2 -
\frac{2\pi\a}{4}f_3(-T)(F^{-})^2 - q_{+}F^{+} - q_{-}F^{-}
\bigg]\,,\label{0AString0}\eea
where $F_{\pm}$ denote field strengths of two RR gauge fields and $q_{\pm}$ denote the corresponding charges, respectively.
The theory admits the following linear dilaton geometry as a vacuum solution:
\bea \Phi &=& -k\varphi\,, \nn \\
ds^2 &=& -dt^2 + d\varphi^2\,,
\label{vacuum}\eea
where all other fields vanish.
 It is also known that there exist charged black hole solutions in the model~\cite{Gukov:2003yp}\cite{Berkovits:2001tg}.
When the tachyon field $T$ is turned off, RR gauge fields can be
easily solved as
\begin{equation}
F^{+}_{01}=F^{-}_{01}=\frac{q}{2\pi\a}\,,\qquad  T=0~,
\end{equation}
which corresponds to the configuration of the background D0-branes
with charges given by $q_{\pm}=q$\,.

One may use this to integrate out RR gauge fields, and obtain the action of the form
\beq I_{0}' = \int d^2 x \sqrt{-g}\,\bigg[\frac{1}{2\kappa^2}~
e^{-2\Phi}\Big(R + 4\nabla_{\mu}\Phi\nabla^{\mu}\Phi + 4k^2  \Big) +\Lambda
\bigg]\,.\label{0AString1}\eeq
Here we denoted the original cosmological constant as $k^2 =
2/\alpha'$ and new effective cosmological constant, coming from
the gauge field contributions, as $\Lambda=-q^2/(2\pi\a)$.
Therefore the low energy effective theory of type 0A string theory
reduces to the two dimensional dilaton gravity with two kinds of
cosmological constants, one of which is related to the charges of
RR gauge fields. The theory admits the charged black hole
solutions in which the dilaton is taken to be proportional to the
spatial coordinate $\varphi$,
\begin{equation}
\Phi_0 = -k\varphi
\end{equation}
while the metric is  of the form
\beq ds^2 = -l_0(\varphi) dt^2 + \frac{d\varphi^2}{l_0(\varphi)} \,,\label{metric}
\eeq
with the factor $l_0(\phi)$ given by\footnote{We set $2\kappa^2=1$ from now on.}
\beq l_0(\varphi) = 1 - e^{-2k\varphi}\,
\frac{1}{2k}\Big(M_0-\Lambda\varphi\Big) \,.
\eeq
As will be clear, $M_0$ may be regarded as a mass of the black
hole. One can also obtain the extremal black hole solution where
the position of  the horizon and the mass are given in terms of
the charge as
\beq e^{2k\varphi_{ex}} = - \frac{\Lambda}{4k^2}\,, \qquad M_{ex}
= \Lambda \,\varphi_{ex} + 2k\,e^{2k\varphi_{ex}} =
-\frac{\Lambda}{2k}\Big[1-\ln\Big(-\frac{\Lambda}{4k^2}\Big)\Big]
\,. \eeq
The near horizon geometry of the extremal black hole becomes
$AdS_2$.

\subsection{The free energy in the 0A matrix model}
In this section we give the expression of free energy in the 0A
matrix model~\cite{Douglas:2003up}\cite{Maldacena:2005he}. The
free energy, $F\equiv\ln Z$, of the 0A matrix model with the Fermi
energy level $\mu$ and Ramond-Ramond(RR) flux $q$ compactified at
the radius $R$ (or at the temperature $1/(2\pi T)=\beta/(2\pi)$)
is given by
\beq \frac{\p^3 F_{0A}}{\p\mu^3} = \Big(\frac{\a}{2}\Big)\, 2R~
{\rm Im} \Bigg[\int^{\infty}_{0} dt \frac{t/2}{\sinh(t/2)}
\frac{\sqrt{\frac{\a}{2}}t/2R}{\sinh(\sqrt{\frac{\a}{2}}t/2R)}~
e^{-i\sqrt{\frac{\a}{2}}\mu t - \half qt}\Bigg]\,. \eeq
In this integral form non-perturbative effects for $R$ and $\a$
are included. Since it is enough for us to consider the
perturbative effects, we perform the series expansion on $x/\sinh
x$ and integrate term by term. This gives us the perturbatively
expanded form as
\bea  \frac{\p^3 F_{0A}}{\p\mu^3} &=& \Big(\frac{\a}{2}\Big)\, 2R~
{\rm Re} \Bigg[\sum_{m=0}^{\infty}(-1)^{m+1}(2m)!
\Big(\mu{\textstyle \sqrt{\frac{\a}{2}}}
- i\frac{q}{2}\Big)^{-2m-1} \times  \nn \\
&& \qquad \qquad \qquad~~~
\sum_{n=0}^{m}\frac{|1-2^{1-2(m-n)}||1-2^{1-2n}||B_{2(m-n)}||B_{2n}|}
{[2(m-n)]!(2n)!}~ \Big(\frac{\sqrt{\a/2}}{R}\Big)^{2n}\Bigg]\,.
\eea
Note that index $m$ corresponds to the genus expansion in the 0A
string theory as $\Big(\mu{\textstyle \sqrt{\frac{\a}{2}}}
- i\frac{q}{2}\Big)^{-1}$ in the matrix model plays the role of the string coupling $g_s$ in the type 0A theory.
On the other hand, the summation over $n$ corresponds to the $\a$-expansion.

Since the thermodynamic free energy, $\CF_{0A}$ is defined by
\beq  F_{0A} \equiv -\beta\CF_{0A}\,, \eeq
we obtain
\bea \CF_{0A} &= &-\sqrt{\frac{2}{\a}} \frac{1}{\pi}~ {\rm
Re}\Bigg[-\half\Big(\mu{\textstyle \sqrt{\frac{\a}{2}}} -
i\frac{q}{2}\Big)^2\ln \Big(\mu{\textstyle \sqrt{\frac{\a}{2}}} -
i\frac{q}{2}\Big) + \frac{1}{24}\,\Big\{1+\Big(
\frac{\sqrt{\a/2}}{R}\Big)^2\Big\}\ln \Big(\mu{\textstyle
\sqrt{\frac{\a}{2}}} - i\frac{q}{2}\Big) \nn
\\ && + \sum_{m=2}^{\infty}(-1)^{m-2}(2m-3)!\Big(\mu{\textstyle
\sqrt{\frac{\a}{2}}} - i\frac{q}{2}\Big)^{2-2m}~ F_{m}(R)\Bigg]\,,
\eea
where
\beq F_{m}(R) \equiv
\sum_{n=0}^{m}\frac{|1-2^{1-2(m-n)}||1-2^{1-2n}||B_{2(m-n)}||B_{2n}|}
{[2(m-n)]!(2n)!}~ \Big(\frac{\sqrt{\a/2}}{R}\Big)^{2n}\,. \eeq

To match the matrix model results with those of the type 0A string
theory on the two dimensional extremal black hole, we should take
some appropriate limits in the above. First of all, we should take
the infinite radius limit({\it i.e.} zero temperature limit ) with
$\mu=0$. Furthermore, we should double the RR flux $q$ to get the
effects from two kinds of RR flux $q_{+}=q_{-}=q$. After all these
limits are taken,  we obtain the tree level part of the
thermodynamic free energy as
\beq  \CF^{tree}_{0A} = -\sqrt{\frac{2}{\a}}\, \frac{1}{\pi}~ {\rm
Re} \Bigg[ -\half \Big(\mu {\textstyle \sqrt{\frac{\a}{2}}} -
iq\Big)^{2}\ln \Big(\mu{\textstyle \sqrt{\frac{\a}{2}}} -
iq\Big)\Bigg]_{\mu=0} = -\sqrt{\frac{2}{\a}}\, \frac{q^2}{4\pi}\ln
q^2\,, \eeq
Note that we have suppressed the ambiguous quadratic terms on $q$. They
 are related to the divergent parts which exist in the free
energy expressions and may be regularized by the explicit cutoff.
One may also note that there is no further $\a$ correction in the
tree level free energy of the type 0A matrix model, which is not
the case for higher genus ones. In the next section, we will
consider higher derivative corrections to the low energy effective
action of type 0A string theory and check that it gives the same
free energy with the type 0A matrix model.

\section{Black holes in the higher derivative type 0A gravity}

Now we would like to include the higher derivative correction,
i.e. higher order $\a$ correction to the  action given in the
previous section. We restrict ourselves to the higher derivative
terms in the NS-NS sector fields only\footnote{One may include the
generic higher derivative terms in the RR sector as well.

}. The results from $\beta$
function computation~\cite{Metsaev:1987zx} tell us that we do not need to consider higher derivative terms in $\Phi$. Furthermore,
the unique higher derivative term for the metric in two dimensions, which appear in the next order in $\a$ correction, is a $R^2$ term.
Henceforth, it is enough to add the following correction term to the action (\ref{0AString1}):
\beq I_1 = \frac{1}{2\kappa^2}~\int d^2 x \sqrt{-g}\,
e^{-2\Phi}
\Big( a\a R^2 \Big)\,,\label{LEEA0AString}\eeq
where  $a$ is a certain dimensionless number
which may be fixed by the computation of $\a$ corrections in string theory.

\subsection{Black hole solutions}

The equations of motion of the metric and the dilaton are found to be
\bea 0&=&R_{\mu\nu} + 2\nabla_{\mu}\nabla_{\nu}\Phi +
g_{\mu\nu}\Big[ 2\nabla^2\Phi - 4(\nabla\Phi)^2 + 4 k^2 + \half
e^{2\Phi}\Lambda\Big]  -\frac{4a}{k^2}\,
e^{2\Phi}\nabla_{\mu}\nabla_{\nu}(e^{-2\Phi}R)\,, ~~~\nn \\
0&=& R + 4\nabla^2\Phi - 4(\nabla\Phi)^2 +  4 k^2 +
\frac{2a}{k^2}\, R^2\,.
\label{EOM}
\eea
The metric for the black hole solutions of the above equations of motion is chosen to be the Schwarzschild type as
\beq ds^2 = -l(\varphi) dt^2 + \frac{d\varphi^2}{l(\varphi)} \,.\label{schwarz}
\eeq
In general, we have three equations of motion in which only two of them are independent.
After simple manipulation, the equations of motion for  the metric component $g_{tt}$ and dilaton $\Phi$ in the above Schwarzschild type metric are given by
\bea 0 &=& l''-6\,l'\, \Phi'-4\, l\,\Phi''+8\, l\, (\Phi')^2-8k^2-\Lambda\, e^{2\Phi}+\frac{4a}{k^2}\,
 (2\,l'\,l''\,\Phi'-l'\,l''') \,, \nn \\
0 &=& l''-4\,l\,\Phi''-4\,l'\,\Phi'+4\,l\,(\Phi')^2-4k^2-\frac{2a}{k^2}\,(l'')^2 \,, \nn \eea
where $\Phi'=\frac{d\Phi}{d\varphi}$ and $l'=\frac{dl}{d\varphi}$. 
It is not easy to obtain a black hole solution analytically.
Instead, we use pertubative approach to find out charged
black hole solutions. We require that the solutions reduce to the
well-known black hole solutions as $a\rightarrow 0$. We also
require that the geometry approaches, asymptotically, to the
linear dilaton geometry (\ref{vacuum}). The word `perturbative'
here means the perturbative expansion in terms of variable `$a$'
in the Eq.~(\ref{LEEA0AString}). Let us recall $a$ is just a
number fixed by $\a$ correction and not a parameter we may
arbitrarily vary. Furthermore there is no guarantee that it is
small. Nevertheless we may still regard it as an adjustable
variable and try to do a perturbative analysis. The partial
justification of this approach is given by the comparison of
results from this approach with those from the exact one in the
near horizon limit of extremal black hole.

First of all, the analysis of the asymptotic behavior of the dilaton and the metric leads to
\bea
\Phi(\varphi) &=& -k\varphi +\CO(e^{-4k\varphi})\,, \nn \\
l(\varphi) &=& 1 - e^{-2k\varphi}\,
\frac{1}{2k}\Big(M-\Lambda\varphi\Big) + \CO(e^{-4k\varphi})\,,
\eea
where we denote `$M$' as a mass of the black hole. As will be shown in later section, there is an ambiguity in defining the ADM mass in two dimensions, due to the divergencies. However the difference $M-M_{ex}$, where $M_{ex}$ is the corresponding quantity in the extremal black hole with the same charge, is well defined as the energy above the extremal black hole, and thus it maybe justified to call $M$ as the mass of the black hole.  
The mass of the black hole depends on $a$ and may be written generically in the form
\[ M = M_0 + aM_1 + a^2M_2 + \cdots\,. \]
We introduce, for clarity, a dimensionless mass $m$ and a dimensionless cosmological constant $\lambda $ as
\[
m=\frac{M}{k}\,, \qquad \lambda =\frac{ \Lambda}{k^2}.
\]

Through the perturbative expansion in $a$, we obtain the solutions of the equations of motion, up to second
order in $a$, as
\bea \Phi &=&  -k\varphi - 16\, a^2\,
e^{-4k\varphi}\Big(m_0 -\lambda k\varphi +\lambda
\Big)^2 + \CO(a^3)\,, \\
l(\varphi) &=& l_0(\varphi) + a\, l_1(\varphi) + a^2l_2(\varphi) +
\CO(a^3)\,, \eea
where
\bea
 l_0(\varphi)&=& 1 +
e^{-2k\varphi}\bigg[-\frac{1}{2}\Big(m_0-\lambda k\varphi\Big)\bigg]\,, \nn \\
l_1(\varphi) &=&e^{-2k\varphi} \bigg[- \frac{m_1}{2}\bigg]+
e^{-4k\varphi}  \bigg[\lambda^2 +
2\Big(m_0
-\lambda k\varphi\Big)^2\bigg]\,, \nn \\
l_2(\varphi) &= & e^{-2k\varphi}\bigg[-\frac{~m_2}{2}\bigg] +
e^{-4k\varphi} \bigg[-96\lambda^2 +
4\Big(m_0-\lambda k\varphi\Big) \Big(m_1 -
48\lambda \Big)
-128\Big(m_0 -\lambda k\varphi\Big)^2\bigg]  ~~~\nn\\
&& +\, e^{-6k\varphi}\bigg[
-8\lambda^3+8\lambda^2
\Big(m_0-\lambda k\varphi\Big) +
48\lambda\Big(m_0-\lambda k\varphi\Big)^2 +
16\Big(m_0-\lambda k\varphi\Big)^3\bigg]\,.
\label{Metric} \eea

At first glance, one may think that $m_i$'s are independent
parameters describing the above perturbative solutions. However,
this is not the case and the mass of a black hole is expressed in terms of just a single
parameter $m$ as can be seen from the fact that all the expressions of
the given solution can be rewritten as the function of a single
variable $m$ up to the given order. In terms of mass parameter $m$, the solution can be rewritten as
\bea \Phi &=&  -k\varphi - 16\, a^2\,
e^{-4k\varphi}\Big(m -\lambda k\varphi +\lambda
\Big)^2 + \CO(a^3)\,, \\
l(\varphi) &=& 1 -
\frac{e^{-2k\varphi}}{2}\Big(m-\lambda k\varphi\Big)+
a e^{-4k\varphi} \bigg[\lambda^2 +
2\Big(m
-\lambda k\varphi\Big)^2\bigg] \nn \\
& &-32 a^2
e^{-4k\varphi} \bigg[3\lambda^2 +
6\lambda\Big(m-\lambda k\varphi\Big)
+4\Big(m -\lambda k\varphi\Big)^2\bigg]  ~~~\label{Metric10} \\
&& +\, 8a^2 e^{-6k\varphi}\bigg[
-\lambda^3+\lambda^2
\Big(m-\lambda k\varphi\Big) +
6\lambda\Big(m-\lambda k\varphi\Big)^2 +
2\Big(m-\lambda k\varphi\Big)^3\bigg] + \CO(a^3)\,.\nn
\eea

This solution has several distinct features. First of all, the only combination which appear in the metric and the dilaton is of the form
\beq
\lambda^p (m-\lambda k \varphi)^q e^{-2(p+q)k\varphi}
\label{form1}
\eeq
Secondly, the coefficients of $a^n$ in the dilaton $\Phi$ include,
generically, terms proportional to $e^{-2lk\varphi}$, with $l=2,
3,\cdots, n$, while the coefficients of $a^n$ in the metric
function $l(\varphi)$  contain terms proportional to
$e^{-2lk\varphi}$, with $l=2, 3, \cdots, n+1$. These generic
features can be shown to be true in all orders of $a$ from the
recursive structure in the equations of motion. From these
properties, we find an additional characteristic feature of the
solution: the solution depends only  on  one
combination ${\tilde m}$ of  two parameters
$m$ and $\lambda$ where
\[
{\tilde m} = \frac{m}{\lambda} -\frac{1}{2}\ln \Big(-\frac{\lambda}{4}\Big)~.
\]
 In order to see this, we introduce the shifted spatial coordinate ${\tilde \varphi}$:
\[
{\tilde \varphi}= \varphi -\frac{1}{2k}\ln \Big(-\frac{\lambda}{4}\Big)~.
\]
Then the generic form displayed in (\ref{form1}) becomes $({\tilde
m}-k {\tilde \varphi})^q e^{-2(p+q)k{\tilde \varphi}}$, which
depends only on the parameter ${\tilde m}$. This property will be
useful later on.

\subsection{Horizon and temperature}

The horizon is determined in terms of given variables $m$ and
$\lambda$ by the condition
\beq l(\varphi_{H}) = 0\,. \label{Horizon}\eeq
Using the perturbative solution of the function $l(\varphi)$ given
in~Eq.(\ref{Metric10}), up to $a^2$ order, one can reorganize the
above relation between the mass parameter $m$ and the horizon $\varphi_{H}$,
which is more suitable to the perturbative determination of
$\varphi_{H}$, as
\[ m - \lambda k\varphi_{H} = 2\, e^{2k\varphi_{H}}\bigg[1 + a\bigg(
8+ \lambda^2\, e^{-4k\varphi_H}\bigg) -
8a^2\,\bigg(32  + 24\lambda\, e^{-2k\varphi_{H}} +
8\lambda^2\, e^{-4k\varphi_{H}}+
\lambda^3\, e^{-6k\varphi_{H}}\bigg)+\CO(a^3)\bigg]\,. \]
The perturbative expression for the position of the horizon, $\varphi_{H}$, can be obtained by expanding it as
a series in $a$ as
\beq \varphi_H =  \varphi^{0}_{H} + a\,\varphi^{1}_{H} + a^2\,
\varphi^{2}_{H} + \CO(a^3)\,,\eeq
which gives us $\varphi^i_{H}$ as functions of $m_i$ $(i=0,1,2 \cdots)$, and, in principle, $\varphi_{H}$ as a function of $m$.

Temperature of these non-extremal black holes can be obtained from the condition of
the absence of the conical singularity in the Euclideanized black
hole geometry. In the case at hand, the temperature is given by the relation
\[ T = \frac{1}{4\pi}~ l'(\varphi_H~ ;~ m,\lambda)\,. \]
Note that in terms of the shifted coordinate, it becomes $ T
=\frac{1}{4\pi}~ \p_{\tilde{\varphi}}l(\tilde{\varphi}_H~ ;~
\tilde{m})$, which tells us that there is no $\lambda$-dependence
in the relation between the Hawking temperature $T$ and the
position of the horizon in the shifted coordinate, ${\tilde
\varphi}_H$.
 As we have seen, $\tilde{\varphi}_{H}$
is determined by the mass parameter $\tilde{m}$
$(\tilde{\varphi}_{H}=\tilde{\varphi}_{H}(\tilde{m}))$ or vice
versa $(\tilde{m}=\tilde{m}(\tilde{\varphi}_H))$ from the
Eq.~(\ref{Horizon}). After expressing the mass parameter
$\tilde{m}$ in the metric function $l(\tilde{\varphi})$ in terms
of the position of the horizon, $\tilde{\varphi}_{H}$, the above
relation reduces to the one between the position of the horizon in
the shifted coordinate and the Hawking temperature. Then, the
formula can be inverted and $\tilde{\varphi}_H$ can be written in
terms of $T$ only.  Up to $a^2$ order the perturbative relation is
given by
\bea  k{\tilde \varphi}_{H}(T) &=& k\varphi_{H}(T)-\frac{1}{2}\ln\Big(-\frac{\lambda}{4}\Big) \nn  \\
&=& -\frac{1}{2}\ln\bigg[1- 2\pi \Big(\frac{T}{k}\Big) - 8a
\bigg\{1 - 4\pi \Big(\frac{T}{k}\Big) + 8\pi^2\,
\Big(\frac{T}{k}\Big)^2\bigg\} +
\CO(a^3)\bigg]\,,\label{HorizonTemp}\eea

One can take the extremal limit of these charged black hole solutions, which is characterized by the conditions
\[ l(\varphi_{ex}) = 0\,, \qquad l'(\varphi_{ex})  = 0\,, \]
where the second condition tells that the Hawking temperature of the extremal black hole is zero.
Up to $a^2$ order, the perturbative solution of the horizon
position is given by
\beq k\varphi_{ex} =  \frac{1}{2}\ln\Big(-
\frac{\lambda}{4}\Big) + 4a+ 16a^2  +
\CO(a^3) \,, \label{extremehorizon}\eeq
while the parameter $m_{ex}$ is given by
\bea m_{ex} =  -\frac{\lambda}{2}\bigg[1-\ln\Big(-
\frac{\lambda}{4 }\Big)\bigg]  -12\lambda\, a+  16\lambda\, a^2 +
\CO(a^3)\,. \eea
Note that the following combination of two parameters depends only on $a$ :
\bea
\frac{1}{\lambda}\Big(m_{ex}- \lambda\, k\varphi_{ex}\Big) =-\frac{1}{2}\Big[1+ 32a  +
\CO(a^3)\Big]
 \,.\eea
This fact can be confirmed to hold for all orders in $a$ by using the equations of motion.

\section{Free Energy and Entropy of Black Holes}

In this section we find the free energy and the entropy of our
black hole solutions using various approaches. First of all we use
Sen's entropy function formalism to obtain the full expression of
entropy of the extremal black hole. Then we use Euclidean action
formalism and Wald's Noether charge method to obtain the free
energy and the entropy of non-extremal black holes.

\subsection{Entropy of extremal black holes}

For the extremal case, the exact expression for the entropy can be
obtained for given Lagrangian by the entropy function formalism of
Sen~\cite{Sen:2005wa}\cite{Sen:2005iz}\cite{Sen:2007qy}\cite{Hyun:2007ii}.
In this section, we use Sen's formalism to obtain the full
expression of the entropy of the extremal black hole. In the next
section, we will obtain the perturbative entropy of non-extremal
black holes and compare it with the results in this section.

One can show that
the following field configurations, which have the $SO(2,1)$ symmetry,
\bea ds^2 &=& v \Big(-x^2dt^2 + \frac{dx^2}{x^2}\Big)\,, \\
\Phi &=& u\,, \eea
are the solutions of the equations of motion (\ref{EOM}). These
configurations may appear as the near-horizon limit of our
extremal black holes. We found that it is the case, up to $a^2$
order, and we believe it holds, generically, all orders in $a$.
This is supported by the fact that our perturbative solutions
interpolate smoothly between the near horizon region, which is
$AdS_2$ geometry, and the asymptotic region, which is linear
dilaton geometry.  It is in contrast to the higher-dimensional
case~\cite{Sen:2004dp}, with higher derivative terms, where it is
not easy to find the solution which interpolates smoothly between
the near horizon and the asymptotic regions.

The entropy function is given by the value of the total Lagrangian at the horizon
\beq F(v,u)
= -2\pi
v\bigg[e^{-2u}\Big(-\frac{2}{v} + \frac{8a}{k^2}\frac{1}{v^2} +
4k^2\Big) + \lambda k^2\bigg]\,. \eeq
The value of $u$ and $v$ are fixed by the extremum conditions:
\[ \frac{\p F}{\p v} =0\,, \qquad \frac{\p F}{\p u} =0 \,, \]
which lead to
\beq v_{*} = \frac{1}{4k^2}\Big(1+\sqrt{1-32a}\Big)\,, \qquad
e^{2u_{*}} = -\frac{2}{\lambda
}\frac{1}{k^2v_{*}}\Big(1-\frac{8a}{k^2v_{*}}\Big)\,. \eeq
The entropy of the extremal black hole is the extremum value of the entropy function and is given by
\beq S = F(v_{*},u_{*}) = - \frac{\pi\lambda}{2}(1+\sqrt{1-32a})\,. \label{allentropy}\eeq
Note that $a$ should satisfy $a\leq \frac{1}{32}$ for the
entropy to be real, which sets the condition for the model to have
an extremal black hole solution. In order to compare with the
perturbative results in the next section,
 we recover the original physical parameter $q^2=-4\pi\lambda$ and expand in $a$
 and obtain
\beq S= \frac{q^2}{4} \Big(1-8a-64a^2\Big) +
\CO(a^3)\,.\label{entropy}\eeq
As will be shown, it is consistent with the results for the
non-extremal black holes in the extremal limit using the Euclidean
action method and the Noether charge one .

\subsection{Non-extremal black holes : Euclidean action approach}
There are several ways to get the entropy and finite temperature
free energy for non-extremal black holes. We employ two well-known
methods to calculate the finite temperature free energy for
charged black holes, namely Euclidean action
approach~\cite{Gibbons:1976ue} and Wald's Noether charge
method~\cite{Wald:1993nt}\cite{Iyer:1994ys}.
In this section, we get the free energy expression for
non-extremal black holes by using the Euclideanized action. Before
proceeding, let us recall that the naive ADM mass is divergent for
a two dimensional black hole which asymptotes to the linear
dilaton background. To regularize this divergence, one should
introduce a cut-off. One way to remove the cut-off dependent terms
is to subtract the value of the reference spacetime. One natural
choice for the reference spacetime is the extremal black hole
which has the same charge in the background as non extremal
ones~\cite{Liebl:1996ti}.

In the Euclidean action approach, we need the
Gibbons-Hawking-York~\cite{York:1972sj}\cite{Gibbons:1976ue}
boundary terms to get the well-defined variational problem and the
correct free energy. Since we have a higher derivative correction
term in the bulk Lagrangian, we should consider the corresponding
correction in the boundary terms, as well. It turns out that
this correction in the boundary terms has no effect to the free
energy expression in the case at hand. The boundary terms, when the bulk Lagrangian is
written in terms of an arbitrary function, $f(R)$, of the Ricci scalar $R$,  can
be easily found and the total action becomes
\beq I_E =-\int_{M} \sqrt{g}f(R) -
2\int_{\p M}\sqrt{h}\, f'(R)\, K\,, \qquad f'(R)
\equiv \frac{d}{d R}f(R)\,,\eeq
where $K$ is the trace of the second fundamental form of the boundary
$\p M$.
The form of the boundary terms is valid even when  $f(R)$ contain other fields, for instance, the dilaton field.
In our case,  $f(R)$ is given by
\[ f(R)= e^{-2\Phi}\Big(R + \frac{2a}{k^2}R^2 + 4\nabla_{\mu}\Phi\nabla^{\mu}\Phi + 4k^2  \Big) + \Lambda \,, \qquad \]
After plugging in our perturbative solution and taking  the boundary at $\varphi = \bar{\varphi}$, the action of the bulk part contributes to
\beq \int_{\varphi_H}^{\bar \varphi} d\varphi \sqrt{g}f(R)  =  4k\, e^{2k\varphi} + 3\lambda
k^2\varphi + aI_1(\varphi) + a^2I_2(\varphi)\,\Big|_{\varphi_H}^{\bar \varphi},  \eeq
where
\bea I_1(\varphi) &=& e^{-2k\varphi}4k\bigg[ \lambda^2 + 2(m-\lambda k\varphi)^2\bigg]\,, \\
I_2(\varphi) &=& e^{-2k\varphi}128k\bigg[-5\lambda^2-
9\lambda(m-\lambda k\varphi) -5(m-\lambda k\varphi)^2
\bigg] \nn\\
&& +\, e^{-4k\varphi}32k\bigg[-\lambda^3 + 5\lambda^2(m-\lambda
k\varphi) + 12\lambda(m-\lambda k\varphi)^2 + 4(m-\lambda
k\varphi)^3\bigg]\,.~~~~~ \eea

The boundary term in the Euclidean action is given by
\beq  \sqrt{h}\, f'(R)\, K \Big|_{\varphi =\bar \varphi}= \frac{1}{2}\, l'(\varphi)\,
e^{-2\Phi}\Big(1+a\frac{4}{k^2}\, R\Big)\Big|_{\varphi =\bar \varphi}\,. \eeq
After plugging in our perturbative solutions,
 the boundary part of the action becomes
\beq I_{bd} = -\beta k\bigg[m - \lambda k\bar{\varphi} +
\frac{\lambda}{2} + \CO(e^{-2k\bar{\varphi}})\bigg]_{\bar{\varphi}\rightarrow\infty}\,. \eeq

The value of the total Euclidean action diverges as the cut-off
${\bar \varphi}$ goes to infinity. We remove these cut-off
dependent terms by subtracting the action value of the reference
extremal black hole with the same charges. The periodicity
$\beta_{ex}$ in the Euclidean time direction for the extremal
black hole is chosen from the
relation~\cite{Gibbons:1994ff}\cite{Hawking:1994ii}\cite{Danielsson:2004xf}
\beq \beta_{ex}\sqrt{g^{ex}_{\tau\tau}(\bar{\varphi})} = \beta \sqrt{g_{\tau\tau}(\bar{\varphi})} \,, \eeq
so that the proper length of the reference geometry in the
Euclidean time direction at the boundary coincides with the one of
the non-extremal black hole geometry. Its asymptotic form becomes
\[ 1- \frac{\beta_{ex}}{\beta} = \frac{M-M_{ex}}{4k}\, e^{-2k\bar{\varphi}}+\CO( e^{-4k\bar{\varphi}}) \,. \]

The regularized free energy is defined by
\[ F_{BH}(\beta) = \frac{I_E(\beta)}{\beta} - \frac{I_E^{ex}(\beta_{ex})}{\beta_{ex}}\,,
\]
which is determined to be, up to second order,
\beq F_{BH}(T) = \Lambda \bigg[\varphi_H(T) -
\varphi_{ex}\bigg]\,, \label{conjecture}\eeq
where $\varphi_H(T)$ and $\varphi_{ex}$ are given in
(\ref{HorizonTemp}) and (\ref{extremehorizon}). Note that the
temperature dependence of the free energy is entirely incorporated
in the horizon position $\varphi_H$ up to the second order in $a$.
We conjecture the relation (\ref{conjecture}) is an exact one,
holding for all orders in $a$. One of the reason that  the
relation (\ref{conjecture}) might be an exact one comes from the
relation, derived in appendix B, between the temperature
dependence of the horizon radius, or the position of the event
horizon in two dimensions, of the non-extremal black hole and the
curvature radius of the $AdS_2$ geometry, which is the
near-horizon geometry of the corresponding extremal black hole.
Indeed, from the relation  (\ref{conjecture}), conjectured to hold all orders in $a$, we obtain the entropy of the
extremal black hole as
\[
S = - \frac{\p F}{\p T}\bigg|_{T=0} = -\Lambda
\frac{\p\varphi_H(T) }{\p T}\bigg|_{T=0} =  - 2\pi \Lambda v_* =
-\frac{\pi \Lambda }{2k^2}\Big(1+\sqrt{1-32a}\Big)
\]
which agrees with the exact expression (\ref{allentropy}) of the entropy of the extremal black hole.

Via a standard
thermodynamic relation, we obtain the entropy of the non-extremal charged black holes as
\bea S &=& - \frac{\p F}{\p T} \\ \nn &=& - \frac{\pi
\Lambda}{k^2} \bigg(1 - 2\pi\Big(\frac{T}{k}\Big)\bigg)^{-1}
\Bigg[1 - 8a \Bigg(\frac{1 - 8\pi(\frac{T}{k}) +
8\pi^2(\frac{T}{k})^2}{1 - 2\pi(\frac{T}{k})}\Bigg) \\ \nn & &
-64a^2 \Bigg(\frac{1 - 12\pi(\frac{T}{k}) + 48\pi^2(\frac{T}{k})^2
- 96\pi^3(\frac{T}{k})^3 + 64\pi^4(\frac{T}{k})^4}{\Big(1 -
2\pi(\frac{T}{k})\Big)^2} \Bigg) \Bigg] + \CO(a^3)\,. \eea
and indeed, in the extremal limit, the entropy becomes
\beq S =
\frac{q^2}{4} \Big(1-8a-64a^2\Big) + \CO(a^3)\,, \eeq
which agrees with the result (\ref{entropy}) from the Sen's formalism in the previous section.

\subsection{Non-extremal black holes : Noether charge method}
Now, we turn to the Noehter charge method pioneered by
Wald~\cite{Wald:1993nt}\cite{Iyer:1994ys}\cite{Iyer:1995kg} and
obtain the same results as Euclideanized action method if we
take the same prescriptions for removing the cut-off dependent
terms. By using this method we may obtain more insights into the
structure of free energy. As was done in Euclidean action
approach, we may regularize the divergent expression by
subtracting the reference spacetime which is given by the extremal
black hole. For the basic materials on the Noether charge method relevant for our discussions, see appendix A. 

The Noether charge for the action variation under the coordinate transformation by vector field $\xi$ is
\beq Q^{\mu\nu}_{\xi}  = -2\bigg[e^{-2\Phi}\Big(1+\frac{4a}{k^2}R\
\Big) \nabla^{[\mu}\xi^{\nu]} + 2\xi^{[\mu}\nabla^{\nu]}\Big(e^{-2\Phi}(1+\frac{4aR}{k^2})\Big)\bigg]\,.   \eeq
The relevant Killing vector field for the ADM-like mass and the free energy is the timelike one $\frac{\p}{\p t}$.
Then, as shown in the appendix A, the energy is given by
\bea \CE &=& Q^{t\varphi} + B^{\varphi}\Big|_{\infty} \eea
where
\bea Q^{t\varphi} &=&
e^{-2\Phi}\Big(1+\frac{4a}{k^2}R\Big)\p_{\varphi}l-2l\p_{\varphi}\Big(1+\frac{4a}{k^2}R\Big)\,,
\\
B^{\varphi} &=&  e^{2k\varphi} \Big[2k\,l-\p_{\varphi}l - 8k\,\Phi
\Big]_\infty \,. \eea
After plugging in the solution, the energy of the charged black hole is found to be
\bea \CE
= M - \Lambda\varphi - 2k\,e^{2k\varphi} + 8k^2\varphi e^{2k\varphi} + \frac{16a\Lambda}{k} \,\Big|_{\infty} \,.\eea
One may note that, as explained in the appendix A, there remain
ambiguities in the choice of surface term $\Theta^\mu$ and,
accordingly, potential for currents $Q^{\mu\nu}$  in the Noether
charge method. In the higher dimensional case, where ADM mass is
well-defined and finite, those ambiguities do not affect the ADM
mass and therefore do not represent any ambiguity. In
two-dimensional case, the ADM-like mass is apparently divergent
and, thus, should be taken with care. One may wonder whether those
apparent ambiguities really lead to any real ambiguity in the
ADM-like mass. It turns out that they affect only on the cut-off
dependent part, which will be removed eventually, leaving the
essential part,
\[ M- \Lambda \varphi + \frac{16a}{k}\Lambda\, \]
invariant.

In the Noether charge method,
the entropy is given by the value of Noether charge at the horizon and thus it  becomes
\beq     S = \frac{1}{T}\, Q^{t\varphi}\Big|_{\varphi_H} = 4\pi~ e^{-2\Phi}\, (1+ \frac{4a}{k^2}\, R)\Big|_{\varphi_H}\,. \eeq
The entropy itself is finite and well-defined. It is totally free
from the ambiguities in defining Noether charge mentioned above.
Through the thermodynamic relation among the free energy, the
energy and the entropy as
\beq  \CF_{BH} =\CE- TS\,, \eeq
we obtain the free energy $F_{BH}$ with respect to the reference geometry:
\bea
 F_{BH} &=& \CF_{BH} - \CF_{BH}^{ex} \nn \\
 &=& (M-M_{ex}) -T \bigg[4\pi ~ e^{-2\Phi}\Big(1+ \frac{4a}{k^2}
R\Big)\bigg]_{\varphi_{H}}\,. \eea
When our black hole solutions are inserted in the above equation, the same
results as Euclidean action approach are obtained, up to the second order in $a$, as
\beq F_{BH}(T) = \Lambda
\bigg[\varphi_H(T)-\varphi_{ex}\bigg] \,. \eeq

So far we have used the extremal black holes as our reference
geometry to remove the cut-off
dependent terms in the mass of
non-extremal black holes. This is the same prescription we used in
the Euclidean action formalism where it seems to be the only
natural choice to deal with cut-off dependent terms. As a result,
the free energy for the extremal black hole is assigned to be
zero. But in the Wald's Noether charge method, there seems to be
more natural prescription in dealing with the cut-off dependent
terms. We may just remove all the cut-off dependent terms in the
mass formula for the non-extremal black hole. If we follow this
prescription, the free energy becomes
\bea
 F_{BH} (T)= (M + \frac{16a}{k}\Lambda) -T \bigg[4\pi ~ e^{-2\Phi}\Big(1+ \frac{4a}{k^2}
R\Big)\bigg]_{\varphi_{H}}\,. \eea
In the extremal limit, the free energy is given by
\bea
 F_{BH}^{ex} = (M_{ex} + \frac{16a}{k}\Lambda) = -\frac{1}{\sqrt{2\a}}\frac{q^2}{4\pi}\ln
q^2 +\CO(q^2)\,. \eea
In the leading term, i.e. $q^2\ln q^2$, there is no
$\a$-correction, which agrees with the result from the matrix
model. There is inherently an overall factor 2 difference between
the free energy of the extremal black hole and the matrix model.
This may be resolved by rescaling the time coordinate of black
hole.

\section{Discussion}

In this paper we constructed charged black hole solutions in the
low energy effective theory of type 0A string theory with $\a$
correction term. Though they are not exact forms, but are given
only perturbatively in $a$, they interpolate smoothly between the
near horizon geometry and the asymptotic dilaton geometry, which
strongly indicates the existence of the solutions. They also have
smooth extremal limit whose near horizon geometry is $AdS_2$,
which can be easily confirmed to be the exact solution of the
equations of motion with the higher derivative term if $a \leq
1/32$.

We computed the free energy and the entropy of the non-extremal
black holes using Euclidean action and Wald's Noether charge
approach. There are divergent or cut-off dependent terms in the
computation of ADM-like mass in two dimensions due to the linear
dilaton. We discussed the subtleties in removing those cut-off
dependent terms. The leading term of free energy in the large
charge limit is independent of the method of removing cut-off
dependent terms and thus shown to be well defined. It was shown to
be the same as the one from matrix model. This supports the
duality between the 0A string theory and the 0A matrix model.
Though the geometry itself is given only perturbatively, we found
the exact expression of the entropy of the extremal black hole and
also found an exact relation between the free energy of the
non-extremal black hole and the the position of Killing horizon.

It is believed that the matrix model does not admit the black hole
solution.  On the other hand there are exact black hole solutions
in the low energy effective theory of the dual type 0A string theory. One argument against the existence
of black hole solutions is that the curvature radius is the order
of string length scale, independent of the charge, and thus low
energy effective gravity theory can not be trusted. We considered
the possible higher derivative term in type 0A gravity and found
the condition of the existence of the black hole solutions. By
direct computations in string theory, one may get the value of $a$
to confirm the (non-)existence of the 0A black holes. 

It would be very interesting to include the higher derivative terms in one-form gauge fields and the combination of them with NS-NS fields. Since the 0A string theory is not supersymmetric, it is not clear how to incorporate those terms. Another interesting problem is whether the duality holds when we include the quantum correction in the 0A string theory due to the tachyon field. Works in this direction is under progress.

{\bf Acknowledgments}

K.~Hur and S.~Hyun were supported by
the Basic Research Program of the Korea Science and Engineering
Foundation under grant number R01-2004-000-10651-0.
The work of S.~Hyun was supported by
 the Science Research Center Program of the
Korea Science and Engineering Foundation through the Center for
Quantum Spacetime \textbf{(CQUeST)} of Sogang University with grant
number R11 - 2005 - 021. H.~Kim and S.-H.~Yi were supported by the Korea Research Foundation Grant funded
by Korea Government(MOEHRD, Basic Reasearch Promotion Fund)
(KRF-2005-070-C00030).

\section*{Appendix A}
In this appendix, we summarize some relevant results in Noether
charge method for the entropy and the free energy by
Wald~\cite{Wald:1993nt}\cite{Iyer:1994ys}
\cite{Iyer:1995kg}\cite{Jacobson:1993vj}\cite{Jacobson:1994qe}.

Under a general field variation $\delta \phi^a $, the variation of the
Lagrangian is given by
\[ \delta (\sqrt{-g}\CL) = \sqrt{-g}E_a\delta \phi^a +
\p_{\mu}(\sqrt{-g}\Theta^{\mu})\,, \]
where $E_a$ denotes the equations of motion for various fields,
$\phi^a$, and $\Theta$ corresponds to the surface term. The Noether current of a vector field $\xi$ for
diffeomorphism invariance can be obtained as
\beq J^{\mu}_{\xi} = \Theta^{\mu}_{\xi} - \CL~
\xi^{\mu}\,. \eeq
The potential $Q$ for the current is defined by
\beq \nabla_{\nu}Q^{\mu\nu} = J^{\mu}\,. \eeq
Though there are ambiguities in the definition of the potential
$Q$, Wald showed that the black hole entropy and the free energy can be
written unambiguously in terms of $Q$ as
\beq S = \frac{1}{T} \int_{H}d\Sigma_{\mu\nu}~ Q^{\mu\nu}\,,
\qquad F = \CE - TS\,, \eeq
where $H$ denotes bifurcate Killing horizon for Killing vector
$\xi$ and $\CE$ is the ADM-like energy which can also be obtained
by Noether charge method. More explicitly, it was shown that,
whenever one can write
\[ \delta
\int_{\infty}
d\Sigma_{\mu\nu}\Big(B^{\mu}\xi^{\nu}-B^{\nu}\xi^{\mu}\Big) =
\int_{\infty}d\Sigma_{\mu\nu}\Big(\Theta^{\mu}\xi^{\nu}-\Theta^{\nu}\xi^{\mu}\Big)\,,
\]
$\CE$ can be defined by
\beq \CE = \int_{\infty}
d\Sigma_{\mu\nu}~\Big(Q^{\mu\nu}-(B^{\mu}\xi^{\nu}-B^{\nu}\xi^{\mu})\Big)\,. \eeq
Since the ambiguities in $Q^{\mu\nu}$ may play some roles in our case, we explain them  in some detail.  One of them comes from
the total derivative terms added in the Lagrangian, which do not
change the equations of motion
\[ \sqrt{-g}\CL \longrightarrow \sqrt{-g}\CL + \p_{\mu}(\sqrt{-g}
\bfmu^{\mu})\,. \]
This additional term modifies $\Theta$ as
\[ \Theta^{\mu} \longrightarrow \Theta^{\mu} + \delta \bfmu^{\mu} +
\bfmu^{\mu}~ \half g^{\alpha\beta}\delta g_{\alpha\beta}\,.
\]
Note that the transformation of the added term in $\Theta$ under the coordinate
transformation ({\it i.e.} its Lie derivative) is given by
\[ \delta_{\xi}\bfmu^{\mu} + \bfmu^{\mu}\half g^{\alpha\beta}\delta_{\xi}g_{\alpha\beta}= \xi^{\alpha} \nabla_{\alpha}\bfmu^{\mu}
 - \bfmu^{\alpha} \nabla_{\alpha}\xi^{\mu} + \bfmu^{\mu}\nabla_{\alpha}\xi^{\alpha}\,. \]
There is another ambiguity in $\Theta$ from its definition.  Namely, one
can modify $\Theta$ as
\[ \Theta^\mu \longrightarrow \Theta^\mu + \nabla_\nu \bf{Y}^{\nu \mu}\,, \]
where $\bf{Y}^{\mu \nu}$ is an antisymmetric tensor. This leads to
the modified form of the Noether current
\[ J^{\mu} \longrightarrow J^{\mu}  +
\nabla_{\nu}\Big(\xi^{\nu}\bfmu^{\mu} - \xi^{\mu}\bfmu^{\nu}+
\bf{Y}^{\nu \mu} \Big)\,. \]
The Noether charge itself has ambiguity in its definition,
$J^{\mu}=\nabla_{\nu}Q^{\nu\mu}$, up to total derivative term
\[ Q^{\nu\mu}\longrightarrow Q^{\nu\mu} + \nabla_{\rho}\bf{Z}^{\rho\nu\mu}\,, \]
where $\bf{Z}^{\rho\nu\mu}$  is a totally antisymmetric tensor.
Therefore, these ambiguities can modify the Noether charge,
$Q^{\mu\nu}$ as
\beq Q^{\mu\nu}  \longrightarrow Q^{\mu\nu} + \xi^{\nu}\bfmu^{\mu}
- \xi^{\mu}\bfmu^{\nu} + \bf{Y}^{\nu \mu} +
\nabla_{\rho}\bf{Z}^{\rho\mu\nu}\,. \eeq
For completeness, we give an example how the above various tensors
are related to Wald's differential
forms~\cite{Wald:1993nt}\cite{Iyer:1994ys}. The tensor
$\bfmu^{\mu}$ is related to the component of the differential form
$\mu_{\beta_1\cdots\beta_{D-1}}$ in $D$-dimensional spacetime case
as
\[ \bfmu^{\mu} = \frac{1}{(D-1)!}\,
\epsilon^{\mu\beta_1\cdots\beta_{D-1}}\,
\mu_{\beta_1\cdots\beta_{D-1}}\,, \qquad
\sqrt{-g}\epsilon^{01\cdots D-1} = -1\,.
\]

\section*{Appendix B}

In this appendix, we derive a somewhat generic relation between the
temperature dependence of the radius of the time-like Killing horizon for a
non-extremal black hole and the curvature radius of the near horizon geometry in the extremal
limit.

For concreteness, let us take the metric as a Schwarzschild type
as
\[ ds^2 = -l(r)\, dt^2 + \frac{dr^2}{l(r)} + r^2d\Omega^2_{D-2}\,.
\]
Now, let us recall relations leading the horizon and temperature
\[ l(r_H) = 0\,, \qquad T=\frac{1}{4\pi}~ l'(r_H)\,, \qquad\qquad
l'\equiv \frac{d}{d r}l(r)\,. \]
If the radius of the horizon and the mass of non-extremal black
holes are assumed to be written as an analytic function of
temperature, which is not a valid assumption in the case of
Schwarzschild black holes, the generic form of the expansions are
\bea r_H(T, Q^a) &=& r_{0}(Q^a) + f(Q^a)\, T + \CO(T^2)\,, \nn \\
m(T, Q^a) &=& m_0(Q^a) + c(Q^a)~T + \CO(T^2)\,, \nn \eea
where $Q^a$ denote various charges in the given setup. Note that
$r_0$ and $m_0$ are the radius of the horizon and the mass of the
corresponding extremal black hole, respectively, and thus satisfy
\[
 l(r_0;\, m_0)=  l'(r_0; \, m_0)=0\,.
 \]
Then, the temperature relation, when $l'(r_H)$ is expanded near
$r_0$, leads to
\[ 4\pi\, T = \bigg[f(Q)~ l''(r_0\,;\, m_0,Q) + c(Q)\, \frac{\p}{\p
m}\, l'(r_ 0\, ; \, m(T),Q)\Big|_{T=0} \bigg]T + \CO(T^2)\,, \]
where there is no temperature independent term on the right hand side due to the extremal
condition, $ l'(r_0\, ; \, m_0,Q) =0$. On the other hand the horizon condition  becomes
\[ 0 = l(r_H\, ;\, m, Q) =  c(Q)\, \frac{\p}{\p
m}~ l(r_H\,;\, m,Q)\Big|_{T=0} T+ \CO(T^2)\,, \]
which gives the condition
\[ c=0\, \qquad\qquad {\rm as}~~~ \frac{\p }{\p m}~ l \neq 0\,. \]
As a result, we get
\beq \frac{\p r_H(T)}{\p T}\Big|_{T=0} = f(Q) =
\frac{4\pi}{l''(r_0\,;\, m_0,Q)}\,, \eeq
which shows the relation between the temperature dependence of the
 horizon's radius of a non-extremal black hole and the
curvature radius of the near horizon geometry in the corresponding
extremal black hole, since the near horizon geometry in the
extremal limit under the assumption $l''(r_0\, ;\, m_0,Q) \neq 0$
is given by
\[ ds^2_{near} = \frac{2}{l''(r_0\, ; \, m_0,Q)} \Big[ -x^2dt^2 +
\frac{dx^2}{x^2}\Big] + r_0^2d\Omega^2_{D-1}\,, \qquad x\equiv
\half\, l''(r_0\, ; \, m_0,Q)~r\,,
\]
%


\end{document}